\documentstyle[preprint,aps]{revtex}
\begin{document}
\draft
\title{Semiclassical Quark Model Of the Nuclear Matter
Response}
\author{W. Melendez and C. J. Horowitz
\thanks{Electronic communication: charlie@iucf.indiana.edu}}
\address{Nuclear Theory Center, 2401 Milo B.
Sampson Lane, Bloomington, Indiana 47408 USA}

\date{\today}
\maketitle

\begin{abstract}
The longitudinal response function $R(q,w)$\ of nuclear matter is
calculated in a semiclassical quark model.  The model has a many-
body string-flip potential that confines quarks into hadrons and
avoids color van der Waals forces.  Molecular dynamics
simulations are used to calculate $R(q,w)$\ in one space dimension
for a variety of momentum transfers $q$\ and excitation energies
$w$.  The response function decreases with density (compared to a
free hadron response) because of quark exchange between hadrons.
\end{abstract}


\section{Introduction}

The response of nuclei to a variety of electroweak and hadronic
probes is rich in excitations involving many nucleons, such
as giant resonances, and in internal excitations of single baryons.
People have speculated on possible mixing between  the ``nucleon-
like'' Gamow-Teller resonance and the ``quark-like'' $N-\Delta$
transition\cite{Delta}.  We propose to study this mixing by
calculating a variety of response functions using a  quark model of
nuclear matter.  Indeed, the string-flip or quark exchange model  of Horowitz
et al.\cite{HMN} and Lenz et al.\cite{Metal} contains both types
of modes on an equal footing.


The nuclear excitation spectrum contains many
modes with different spin, flavor and color quantum
numbers.  One can speculate on new types of modes
involving collective excitations of quarks ``from'' several
nucleons.  For example, there could be a collective oscillation of
up quarks against down quarks (independent of weather
a given quark is in a proton or neutron).  This would be qualitatively
different from both conventional giant resonances  (where, for
example, neutrons oscillate against protons) and
quasifree Delta production (which only involves quarks
``from'' a single hadron).  We call such many-quark
collective modes {\sl quark giant resonances} and will
search for them in our quark model response functions.

We use a string-flip potential model\cite{HMN,Metal} which (1)
confines quarks into hadrons, (2) has the correct
exchange symmetry among identical  quarks (even for
quarks ``from'' different hadrons) and (3) has no color
Van der Waals forces between color singlet
hadrons.\footnote{Such Van der Waals forces
are not observed and are  present in confining
two-body potentials.  They produce large  spurious
energies in nuclear matter.}

The ground state properties of both one
dimensional\cite{HMN,HPPRC} and
three-dimensional\cite{HPNPA,Watson} models of nuclear matter have
been calculated  using variational and path integral (Green's
function) Monte Carlo  techniques.  At low density the model
reduces to a gas of weakly interacting hadrons with properties
similar to those in free space, while at high density the
model reduces to a free Fermi gas of quarks.

A first attempt was made\cite{CJH} to examine the
longitudinal response in the model by calculating the energy
integral of the response function.  This integral or Coulomb sum
can be expressed in terms of the ground
state correlation function.  The latter was directly calculated
via Monte Carlo techniques.  However, it is difficult to interpret
the quark model Coulomb sum since it necessarily has
contributions from both nucleon excitations (such as
the quasielastic peak) and quark excitations of a single
hadron.  To  make further progress, it is desirable to
calculate the distribution of strength in order to
separate the nucleon excitations at low energy from
the  ``quark'' excitations at higher energies.

Unfortunately, it is not possible to directly calculate
the response  function in real time via path integral
Monte Carlo techniques.  This is because the complex
phases for real time evolution introduce  prohibitive
statistical noise.  Some progress has been
made calculating response functions in imaginary
time (see for example\cite{Carlson}).  However,
the imaginary time response can not be directly transformed to real
time because of statistical noise.  Thus, there are
ambiguities comparing  calculations in imaginary time
with the physical real time response function.

Alternatively we consider a (semi) classical
approximation to our quark model.  This allows us to
calculate the real time evolution directly using  molecular
dynamics techniques.  From molecular dynamics
trajectories, the response function can be calculated in
a straight-forward and  unambiguous way.

For our harmonic oscillator string-flip potential
(discussed in section II) many of the full quantum
results are reproduced in a classical  approximation.
For example, the excitation spectrum (multiples of
$\hbar \omega_0$) is unchanged.  Furthermore, the ground state
correlation function $g(r)$\ is approximately reproduced.  We do
this by choosing a fictitious temperature so that thermal
motion simulates some of the zero point motion of the
$T=0$\ quantum system.  [Note our calculations are
fully classical except for this fake temperature.]

Our classical approximation may allow insight into
any collective modes we observe.  New modes
with a classical analog may be simple to  understand.
We expect our classical approximation to be better at
describing long wavelength modes such as zero sound
rather than short wavelength modes such as the
quasielastic peak.  Therefore, we will focus on the long
wavelength response.

In this first study  we consider a very simple model in
one spatial dimension without explicit spin, color or
flavor degrees of freedom.   This same model was
considered in Ref.\cite{HMN}. It is straightforward to
extend our calculations to three dimensions (see for
example \cite{HPNPA}). Furthermore, one can include the
internal degrees of freedom and add additional
two-body interactions in order to reproduce the two
nucleon phase shifts \cite{Oka}.  We hope to extend our
simple model in later work.

Our model is described in section II along with the
molecular dynamics formalism.   Section III describes
the free quark and free hadron response functions.
Next, results for nuclear response
functions at various densities, momentum transfers
and excitation energies are presented in section IV.
Finally, we summarize in Section V.

\section{FORMALISM}

In this section we describe the string-flip potential
model and then the molecular dynamics formalism for
calculating the response function.  Our model has two-quark
hadrons.  Although simple, these have interesting composite
features.  It is straightforward to extend our calculations to
three quark hadrons\cite{HPNPA}.

Our classical approximation does not enforce
antisymmetry in the spatial quark wave function.
However, the relatively large spin-flavor and color
degeneracy of nuclear matter suggests that most of the
time the quarks are antisymmetrized in the internal
coordinates.  Therefore, the wave function is symmetric in the spatial
coordinates.
We hope to improve the treatment of internal degrees
of freedom in future work.  For now, we consider the
simplest model without explicit spin, flavor, or color
degrees of freedom.

\subsection{String Flip Model}

The Hamiltonian of our model is given by (for a one
dimensional system),

\begin{equation}
H = T + V,
\end{equation}
where,
\begin{equation}
T= \frac{-\hbar^{2}}{2m} \sum_{i=1}^{N}
\frac{\partial^{2}}{\partial x_{i}^2},
\end{equation}

\begin{equation}
V = {\rm min}_{\{ P\}} \sum_{n=1}^{N/2}
v(x_{P(2n-1)}-x_{P(2n)}). \end{equation}
Here the minimum is taken over all permutations
$P$\ of the quark labels which pair the N quarks into
N/2 pairs, with the confining potential $v$\ or link
acting only between the two quarks in each of the N/2
pairs.   This insures that all quarks
will be confined into hadrons and that there will be no
color Van der Waals forces between hadrons.
Furthermore, the minimum prescription insures
the potential is symmetric in all quarks coordinates.
For the four quark system (for example), the potential
is,

\begin{equation}
V(x_1,x_2,x_3,x_4)={\rm min}\{ V_1,V_2,V_3 \},
\end{equation}

\begin{equation}
V_1=v(x_{12})+v(x_{34}),
\end{equation}
\begin{equation}
V_2=v(x_{13})+v(x_{24}),
\end{equation}

\begin{equation}
V_3=v(x_{14})+v(x_{23}),
\end{equation}
Thus, although the potential is an N-body interaction (i.e.,
it depends upon the configuration of all N quarks), it is
basically quite simple.  The confining forces operate only
between pairs of quarks.  At any instant, the $N/2$\
links arrange themselves in a way that minimizes the
potential energy; this idea is familiar in adiabatic
approximations to field theories.  The only hadron
interactions come from the exchange of quarks
between clusters.

The minimum pairing for a large N system (in three space
dimensions) can be determined with special techniques (see for
example \cite{HPNPA} and \cite{Econ}).  However, in one
dimension  (assuming periodic boundary conditions)
this problem is simple.  Simply order the coordinates
from left to right and then the minimum is one of only
two possibilities,

\begin{equation}
V= {\rm min}(V_R,V_L),
\end{equation}
where
\begin{equation}
V_R = \frac{1}{2} \sum_{i=odd} (x_{i}-x_{i+1})^{2},
\end{equation}

\begin{equation}
V_L = \frac{1}{2} \sum_{i=even} (x_{i}-x_{i+1})^{2}  +
\frac{1}{2}  (x_{N}-x_{1}-L)^{2}.
\end{equation}
Here $L$\ represents the length of the box.
Any other pairing will involve overlapping links and have higher energy.
Note, for simplicity we have assumed harmonic links,
\begin{equation}
v(r)=({k\over 2}){r^2}.
\end{equation}
We discuss below our choice for the oscillator constant $k$.

This model Hamiltonian has been used in previous
works for quantum mechanical calculations
\cite{HMN,Metal,HPPRC,HPNPA,Watson,CJH}. The
expectation value of the energy E is calculated for a
trial wave function   (note as discussed before, we are assuming
Bose symmetry for the spatial coordinates),
\begin{equation}
\Psi = \exp\{-\lambda V\}.
\end{equation}
Here $V$\ is the full many-body potential and
$\lambda$\ is a variational parameter. The value of
$\lambda$ is chosen such that $E =
\langle\Psi|H|\Psi\rangle$\ is a
minimum for a given density $\rho$ of our system. At
low densities, $\lambda = 1/\sqrt{2}$\ reproduces the
oscillator wave-function for isolated hadrons while at
high densities $\lambda \rightarrow 0$\ reproduces
a free particle system. It is
found that as the density increases, the value of
$\lambda$\ that minimizes the energy decreases. The
value of $\lambda$\ for each density will be used to
determine the temperature of our many-body system
for the molecular dynamic simulation (see below).

Our results can be readily scaled for different values of
$m$\ and $k$.  For convenience we chose
the length and energy scales such that the quark mass
$m$, the spring constant $k$\ and the Planck's
constant $\hbar$\ are all equal to one.  The energy
scale is determined by the energy of a free
hadron and is given by
\begin{equation}
E_{0} = (n + \frac{1}{2})\hbar\omega,
\end{equation}
with
\begin{equation}
\hbar\omega = \hbar \sqrt{\frac{2 k}{m}}.
\end{equation}
The length scale is determined by the root
mean square seperation of the quarks in the ground state hadron and is
given by,
\begin{equation}
\langle x^{2} \rangle = \int_{-\infty}^{\infty} \rho(x)
x^{2} dx,
\end{equation}
where
\begin{equation}
\rho(x) = (\frac{mk}{2 \hbar^{2} \pi^{2}})^{1/4}
\exp\{-(\frac{mk}{2 \hbar^{2}})^{1/2} x^{2}\}.
\end{equation}
After evaluating the integral we get,
\begin{equation}
\langle x^{2} \rangle ^{1/2} =
(\frac{\hbar^{2}}{2mk})^{1/4} = 0.841 \
(\frac{\hbar^{2}}{mk})^{1/4}.
\end{equation}

These are our scales for energy and length in our
calculation.  They might be set to describe the radius
of a nucleon, or a fraction of the charge radius
described by a quark core (neglecting the pion cloud)
and a typical baryon excitation scale.

We determine $\lambda$\ by minimizing the
expectation value of the energy $E$.  Note that
\begin{equation}
E = (2\lambda^{2} + 1)\langle V\rangle_{\lambda},
\end{equation}
where
\begin{equation}
\langle V\rangle_{\lambda} = \frac{\int V
e^{-2\lambda V}d\tau}{\int e^{-2\lambda V}d\tau}.
\end{equation}
The equation for $E$\ was obtained by using Eqs.
(1,2,8-10,12). In principle, one  could proceed and calculate
$\frac{dE}{d\lambda}$ , setting the resulting
expression equal to zero and determining $\lambda$\
numerically such that the  derivative is zero. But there
is a simpler way of doing this using the following scaling
property\cite{HPNPA}:
\begin{equation}
\langle V\rangle_{\rho , \lambda} =
(\frac{\rho'}{\rho})^{2} \langle V\rangle_{\rho' ,
\lambda'},
\end{equation}
This property is used to get expressions for $\lambda'$
, $\rho'$ and  $E(\rho')$ in terms of two input
parameters which we denote by $\lambda$ and
$\rho$.

Summarizing, the prescription is as follows : give
$\lambda$\ and $\rho$\ as  two input values. Then
calculate $\lambda'$, $\rho'$\ and $E(\rho')$ using the
equations given in ref\cite{HPNPA}. The value of $\lambda$ that
minimizes the energy $E$\  is our $\lambda'$.

\subsection{Molecular Dynamics Simulation}

A molecular dynamics simulation determines the
motions of the particles in a system\cite{Haile}. It is based on
the microcanonical ensemble and assumes ergodicity.  Our
many-body potential, eq. (3) or (8), is used in conjunction with
Newton's equations of motion to determine the
trajectories of the particles.

The steps of our molecular dynamics simulation are
the following :

a) The initial coordinates of the particles are determined by
using the partition function in a Metropolis simulation.
The partition function is given by,
\begin{equation}
Z = {\rm exp}\{ -V/k_{B}T \} ,
\end{equation}
where $k_{B}$ is the Boltzmann constant and $T$ is
the temperature of the system. The temperature of the
system is determined by the prescription,
\begin{equation}
k_{B}T = \frac{1}{2\lambda} ,
\end{equation}
where $\lambda$\ was determined in
a quantum variational calculation (see fig. 1a).  This prescription
for choosing a fake temperature insures that the
classical calculation will reproduce the correlation
function $g(r)$\  and other ground state observables of
the variational quantum mechanical calculation.  That is,
instead of sampling the  square of the wave  function
eq. (12) we sample an equivalent partition function eq.
(21) in the classical calculation.
This partition function is used in a succession of
Metropolis sweeps to evolve the system  until it
reaches equilibrium . We used 1000 Metropolis sweeps
to determine the initial coordinates. This procedure
simulates quantum zero point motion with thermal
motion.

b) The initial velocities of the particles
are randomly selected from the Boltzmann distribution
given by,

\begin{equation}
f(v) = \sqrt{\frac{2\pi k_{B}T}{m}}
\exp\{\frac{-\frac{1}{2} m v^{2}}{k_{B}T}\}.
\end{equation}

c) The trajectories of the particles are determined by
numerically integrating Newton's equations of motion
for which we used an algorithm by Beeman\cite{Bee}.
Here a time step of 0.01 insures that energy is
conserved to better than one percent (see fig. 1c).

d) From the trajectories we can calculate any time-dependent
quantity of interest. The time-dependent
quantities considered here are the following :

(i) One-body density,
\begin{equation}
Q(q,t) = \frac{1}{\sqrt{N}} \sum_{j=1}^{N} e^{iq
x_{j}(t)}
\end{equation}
This represents the Fourier transform of the charge
distribution of the system where all the quarks are
assumed to be point-like, with unit charge.  In a later work
we will consider quarks of different flavors and
charges.

(ii) Two-body density or radial distribution function,
\begin{equation}
\langle g(r)\rangle_{\rm time},
\end{equation}
where
\begin{equation}
g(r) = \frac{1}{\rho (N-1)} \sum_{i<j} \delta (r - |r_{i}
- r_{j}|).
\end{equation}
Here $\rho$\ is the density of the system, $N$\ is the
number of particles and $r_{i}$\ and $r_{j}$\ are the ith
and jth particle positions , respectively.  This
correlation function is proportional to the probability of
finding a particle at a distance r from another given
particle and is normalized to one for large r.

(iii) Autocorrelation function,
\begin{equation}
F(q,t) = \langle\frac{1}{\tau} \int_{0}^{\tau} Q(q,s)\
Q^{*}(q,s+t)\ ds\rangle.
\end{equation}
The brackets mean that we take an ensemble average, that is,
an average over many different initial conditions.
For $t=0$\ $F(q,0) = S(q)$\  where
$S(q)$\ is the static structure factor.  $S(q)$
can also be expressed as,
\begin{equation}
S(q) = 1 + \frac{N}{L} \int\ dr\ e^{iqr} \ [g(r) - 1].
\end{equation}
{}From eq. (27), one can show that,
\begin{equation}
F(q,-t) = F^{*}(q,t),
\end{equation}
which will be useful in simplifying the calculation for
the response function.

(iv) Response function or dynamical structure factor,
\begin{equation}
R(q,w) = \frac{1}{2\pi} \int_{-\infty}^{\infty} dt\
e^{iwt}\ F(q,t).
\end{equation}
By using the property $F(q,-t) = F^{*}(q,t)$ we can
rewrite $R(q,w)$ as,
\begin{equation}
R(q,w) = \frac{1}{\pi} {\rm Re}
\int_{0}^{\infty}dte^{iwt}F(q,t).
\end{equation}
Thus $R(q,w)$\ is real.
Of course the integral of $R(q,w)$\ gives  $S(q)$,
\begin{equation}
S(q) = \int_{-\infty}^{\infty} dw\ R(q,w).
\end{equation}
The response function is the goal of our calculation
since it will provide us with information about the
excited states and collective modes of the many quark
system.

\subsection{Autocorrelation formalism}

The longitudinal response function $R(q,w)$ describes the
response of a system to a weak external probe.   In
inelastic electron scattering $R(q,w)$\ is proportional
to the differential cross section $d\sigma$\cite{FW}. An
electron probe will interact with the charge density
of the target nucleus (we ignore the transverse response
for now).  Therefore, the interaction
Hamiltonian, $H_{int}$, will be proportional to the
charge density, $\rho(x)$. The charge density is given
by,
\begin{equation}
\rho(x) = \sum_{i}^{N} e_{i} \delta(x-x_{i})
\end{equation}
where for our calculation $e_{i} \equiv 1$.
One can show (by using Fermi's golden rule) that
$d\sigma \propto R(q,w)$ where,
\begin{equation}
R(q,w) = \sum_{f} |\langle f|\rho_{q}|i\rangle|^{2}
\delta(E_{f} - E_{i} - w),
\end{equation}
is the quantum mechanical definition for the response
function. Here $\rho_q = \sum_{j} e_{j} e^{iqx_{j}} \equiv
Q$ is the Fourier transform of the charge density
$\rho(x)$.

The quantum mechanical $R(q,w)$\ can be rewritten
as,
\begin{equation}
R(q,w) = \int_{-\infty}^{\infty} \frac{dt}{2\pi} e^{iwt}
\langle i|\rho_{q}^{\dagger}(t) \rho_{q}(0)|i\rangle.
\end{equation}
Now, in our case we are dealing  with a classical
system, so $R(q,w)$ becomes,
\begin{equation}
R(q,w) = \int_{-\infty}^{\infty} \frac{dt}{2\pi} e^{iwt}
\langle\frac{1}{\tau} \int_{0}^{\tau} dt'
\rho_{q}^{*}(t+t') \rho_{q}(t')\rangle .
\end{equation}
Thus we are calculating in eqs. (27,31) the classical analog of the
quantum mechanical response function.

\section{Free quark and free hadron results}

We will compare our nuclear matter results to the
response of a gas of free quarks and to the response of a
collection of free hadrons.
Therefore  in this section we derive analytic expressions for these two
responses.

\subsection{Free Quark Response Function}

The trajectories of a free quark system with no
interactions are straight lines and consequently we can
find a closed expression for the response function.
Suppose we have $N$\ quarks, where each of them has
a given initial velocity $v_{oi}$ and a given initial
position $x_{oi}$. Then, the trajectories are,
\begin{equation}
x_{i}(t) = x_{oi} + v_{oi} t.
\end{equation}
This implies that $x_{i}(s+t) = x_{i}(s) + v_{oi}t$ which
then gives us, using eq. (24),
\begin{equation}
Q(q,s)Q^{*}(q,s+t) = \frac{1}{N} \sum_{i} \sum_{j}
e^{iqx_{i}(s)} e^{-iqx_{j}(s+t)}.
\end{equation}
Now, when we integrate over the time variable s in
order to calculate the autocorrelation function, only the
case $i=j$\ gives a contribution. Therefore,
\begin{equation}
F(q,t) = \frac{1}{N} \sum_{j} e^{-iqv_{j}t}.
\end{equation}
By using our expression for $R(q,w)$ we get that,
\begin{equation}
R(q,w) = \frac{1}{N} \sum_{j} \delta(w-qv_{j}).
\end{equation}
We take the average of this expression for a
Boltzmann velocity distribution to give,
\begin{equation}
R(q,w) = \int_{-\infty}^{\infty} dv \delta(w-qv)
\frac{1}{\sqrt{2\pi k_{B} T}} e^{-v^{2}/(2 k_{B} T)}.
\end{equation}
Performing the integration we finally get\cite{Hansen},
\begin{equation}
R(q,w) = \frac{1}{\sqrt{2\pi q^{2} k_{B} T}}
e^{-w^{2}/(2 q^{2} k_{B} T)}.
\end{equation}
This is our expression for the response function of a
free quark system. To get the autocorrelation function
$F(q,t)$\ simply Fourier transform eq. (42),
\begin{equation}
F(q,t) = \exp\{-\frac{k_{B} T q^{2} t^{2}}{2}\}.
\end{equation}

\subsection{Free Hadron Response Function}

The free hadron case implies that two quarks are
interacting via a harmonic oscillator potential and
there is no exchange of partners.  For this case we
know the exact trajectories for each of the two quarks and can
calculate the response as shown in the appendix.

\section{Results}

We now present and discuss the results obtained in our
calculation. In Fig 1b we plot sample trajectories
for $N=8$\ quarks as a function of time for a  density of
$\rho=0.414$.  Note that periodic boundary conditions are used with a
box length of $L=19.32$\ (which is $N/\rho$).  This
figure illustrates many of the features of the model.

At t=0 the 8 quarks are grouped into four separate hadrons by the
string-flip potential.  For an isolated hadron, the potential is
simply a harmonic oscillator, so that the
trajectories consist of two curved lines oscillating
together.  The average slope of the two lines gives the
center of mass velocity of the hadron and the amplitude of the
oscillations gives the internal excitation energy.
Near x=0 and t=32 we see a hadron move through the left
wall of the box.  Since we are using periodic boundary
conditions the two quarks reappear near x=19 at the far
wall of the box.

A variety of different kinds of hadron-hadron collisions
are possible with our potential via string rearrangement.
For example, near x=5 and t=23 there is an inelastic
collision between a rapidly downward moving hadron and a
slower upward moving hadron.  After the collision, the
hadron which emerges at larger x has a larger internal
excitation energy which is indicated by the larger
oscillation amplitude.  Clearly this energy came from the
center of mass motion of the two original hadrons.
Note that hadrons only interact via string rearrangement.  If
two quarks are very close together, it is likely they will
remain paired so that the probability to interact is small.
For example, near x=5 and t=87 there is a hadron moving
upwards with the two quarks very close together.  As
result the quarks are only slightly perturbed when this
hadron passes through a larger downward moving hadron.
This behavior follows from our assumption about the
saturating nature of the forces (once two quarks are
paired off the remaining interactions are zero).
It may be the analog, in our very simple model world, of
QCD inspired ideas regarding color transparency.

In Fig.1c we also show the total energy of the N=8 quark
system vs. time.  Our integration method with a step size of dt=0.01
conserves energy to better than one percent.
Furthermore, there is little evidence of a long term gain
or loss in energy.  The hardest part of determining the
trajectories is finding when the strings flip.  We simply
check for flipping at each time step.  No attempt has been made to
try and evolve the harmonic oscillators analytically
between string-flips.

In fig. (2) we show the radial distribution function
$g(r)$\ calculated in two completely different ways. The
large value of $g(r)$\ for small r represents the other quark
which is bound into a given hadron by the string-flip
interaction.  The  dots refer to a Metropolis
calculation which could either represent a quantum Monte
Carlo sampling of the variational wave function of eq. (12)
or a sampling of the classical partition function of eq.
(21).   The stars refer to a Molecular Dynamics simulation.
We get good agreement between the two methods.  This
provides a check of our Molecular Dynamics
numerics and illustrates our simulation of quantum zero
point motion with a fictitious temperature.
In Fig. (3) we show the
autocorrelation function $F(q,t)$\ as a function of time.
The curve with the stars corresponds to a free hadron
calculation, the curve with the
dots corresponds to the string-flip potential and
the solid curve corresponds to the free quark case.
We recall that from the definition of $F(q,t)$, this
function measures how the value of $Q$\ at $s+t$\ is
correlated with its value at $t$. If particles are moving
very slowly (low velocities), then $F(q,t)$\ will fall
off, as $t \rightarrow \infty$, very slowly since the
value of $Q$ at $s+t$ is very correlated to its value at
$t$. The opposite will happen if the particles are moving
very rapidly.  We see that
the free quark $F(q,t)$ falls off faster than
the other two autocorrelation functions. The single
hadron $F(q,t)$ is generally above the corresponding
curves for the free quark and the string
flip potential. We note that the string-flip $F(q,t)$\
oscillates slightly negative before going to zero at large t.   The
response function $R(q,w)$\ is shown in Fig. (4) for a low density.
At this low density, the string-flip
response function is very close to the response for free
hadrons.  As the density is increased in Figs. (5-8) we
see that the string-flip response becomes smaller than the free
hadron result (for low excitation energy w) and
eventually approaches the free quark response at high
densities.

This decrease in the string-flip response with density
illustrates the decrease in the coherence of scattering from
the two quarks in a hadron (without break-up).  At low density and momentum
transfer one scatters coherently from both quarks.
However, as the density increases, it becomes possible for quarks
to hop from one hadron to another because of the
string-flip potential.  This suggests that quark exchange
effects will decrease the response of nuclear matter at
low excitation energies.  We will discuss this more in the
next section.

In Fig. (6)  we show the response function $R(q,w)$\ for
different momentum transfers $q$. Results at high q look
qualitatively similar to the low q responses. This may be an artifact of
our semiclassical approximation.  We expect the classical results to be better
at
low q.

Finally, in Figs
(9-10) we test for finite size effects by comparing
response functions calculated with N=8 and N=16 quarks.    At a low
density of $\rho=0.211$, finite size effects
are very small.  They are somewhat larger at $\rho=0.414$,
but they are still small.  Therefore we do not expect finite
size effects to change our N=8 results greatly.

\section{Summary and Conclusions}

In this paper we have presented a semiclassical
simulation of a simple quark model of nuclear matter.  We
have used thermal motions from a fictitious heat bath to
simulate quantum zero point motion.  Our string-flip
potential model: (1) confines quarks into hadrons, (2)
allows the hadrons to separate without color Van der
Waals forces and (3) is symmetric in all of the quark
coordinates.  The low density limit of the model is a gas
of free hadrons, while the high density limit is a free
quark gas.  This string-flip model has been used
previously for a variety of quantum calculations of
nuclear matter ground state properties.
We have calculated the longitudinal response function from
a molecular dynamics simulation.  We find that $R(q,w)$\
progresses smoothly from that for free hadrons to free
quarks as the density increases.  Compared to the response of free
hadrons, $R(q,w)$\ is found to decrease with
increasing density (at low $w$).  This represents a decrease in
coherence as quarks ``hop" from one hadron to the next, i.e., at
low momentum transfers one scatters coherently
from all of the quarks in a free hadron.  However, the
possibility of quark exchange between nucleons decreases
this coherence in nuclear matter.  Our results suggest
that this quark exchange effect may somewhat decrease the
strength of traditional hadronic calculations of the nuclear response.

Indeed, people have speculated that quark effects may
decrease the Gamow Teller strength, while the experimental
integral of the longitudinal response (the so called Coulomb
sum) could also be below theoretical expectations.
It would be interesting to explore
this suggestion in more sophisticated quark models.

Our simulation provides no evidence for possible new
collective modes.  However, the simulation has been for
the simplest model without explicit spin, flavor, or
color internal degrees of freedom.  In future work, we
will include more internal degrees of freedom (see for
example \cite{GH}) and this will allow us to search a
much richer excitation spectrum for new collective modes
with various spin, flavor, and color quantum numbers.

For simplicity, our calculations have been in one space
dimension.  However, three dimensional simulations should
be straightforward (although requiring more computer
time).  Efficient algorithms do exist for evaluating
the string-flip potential in three dimensions \cite{HPNPA,Econ}.
Finally, we plan to carry out quantum mechanical
calculations of the response function in imaginary time
via Green's function Monte Carlo to compare with these
semiclassical results.

\acknowledgements

Supported in part by Department of Energy grant
DE--FG02--87ER--40365.

\appendix
\section{Free Hadron response function}
In this appendix we calculate the response function for two
isolated quarks interacting via a harmonic oscilator potental
Let the coordinates be $x_{1}$ and
$x_{2}$.  The potential energy between the
two quarks is given by,
\begin{equation}
V = \frac{1}{2} (x_{1} - x_{2})^{2}.
\end{equation}
Introducing  a relative coordinate $x$ and a center
mass coordinate $x_{cm}$\
we can write $x_{1}$ and $x_{2}$ as,
\begin{eqnarray}
x_{1} & = & x_{cm} - \frac{1}{2} x \\
x_{2} & = & x_{cm} + \frac{1}{2} x.
\end{eqnarray}
The trajectories $x_{cm}(t)$\ and $x(t)$\ are given by,
\begin{eqnarray}
x_{cm}(t) & = & A_{cm} + B_{cm}t  \\
x(t) & = & A\ {\rm cos}(\sqrt{2} t + \delta).
\end{eqnarray}
where $A_{cm}$ , $B_{cm}$ , $A$ and $\delta$ are
constants. Substituting these results in the expression
for $F(q,t)$\ and after some simplification and a change
of variable we get,
\begin{equation}
F(q,t) = \frac{2}{\pi} \langle e^{-iqv_{cm}t}
\int_{0}^{\pi} cos[A\frac{q}{2} cos(y)]
cos[A\frac{q}{2} cos(y + \sqrt{2} t)] dy \rangle.
\end{equation}
We can simplify this even more by defining the
function $g(t)$ as,
\begin{equation}
g(t) = \int_{0}^{\pi} cos[A\frac{q}{2} cos(y)]
cos[A\frac{q}{2} cos(y + \sqrt{2} t)] dy,
\end{equation}
were $g(t)$ is periodic with period
$\pi / \sqrt{2}$. Consequently we can
expand $g(t)$ in a Fourier series as follows: let
$-\infty < t < \infty$ with $L = \frac{\pi}{2\sqrt{2}}$ ,
then $g(t+2L) = g(t)$ is $2L$ periodic and therefore,
\begin{equation}
g(t) = A_{0} + \sum_{n=1}^{\infty}(A_{n} cos(\frac{n\pi
t}{L}) + B_{n} sin(\frac{n\pi t}{L})).
\end{equation}
Note $B_{n} = 0$ because $g(t)$ is an even function in
$t$, that is, $g(-t) = g(t)$,so that
\begin{equation}
g(t) = A_{0} + \sum_{n=1}^{\infty} A_{n} cos(\frac{n\pi
t}{L}).
\end{equation}
The coefficients $A_{0}$ and $A_{n}$ are given by,
\begin{eqnarray}
A_{0} & = & \frac{1}{L} \int_{0}^{L} g(t) dt, \\
A_{n} & = & \frac{2}{L} \int_{0}^{L} g(t) cos(\frac{n\pi
t}{L}) dt,
\end{eqnarray}
Inserting these results in the expression for $F(q,t)$
given above we get that,
\begin{equation}
F(q,t) = \frac{2}{\pi} \langle
e^{-iqv_{cm}t}A_{0}\rangle + \frac{2}{\pi}\langle
e^{-iqv_{cm}t}
\sum_{n=1}^{\infty} A_{n} cos(2\sqrt{2} n t) \rangle.
\end{equation}
Recalling that $R(q,w)$\ is the Fourier transform of
$F(q,t)$ we get after substituting eq. (A12) into eq. (30),
\begin{eqnarray}
R(q,w) & = & \frac{2}{\pi} \langle
\delta(w-qv_{cm})A_{0}\rangle + \nonumber \\
& &
\frac{1}{\pi} \sum_{n=1}^{\infty}\{\langle
A_{n}\delta (w+2\sqrt{2}n-qv_{cm})\rangle + \langle
A_{n}\delta (w-2\sqrt{2}n-qv_{cm})\rangle \}.
\end{eqnarray}

Let's write explicitly what the term in brackets means,
so that we can see how this expression is calculated,
\begin{equation}
\langle \delta(w-qv_{cm}) A_{0}\rangle =
\frac{\int_{-\infty}^{\infty}dv_{cm}
\int_{-\infty}^{\infty}dA e^{-E/(k_{B}T)}
\delta(w-qv_{cm})
A_{0}(q)}{\int_{-\infty}^{\infty}dv_{cm}
\int_{-\infty}^{\infty}dA  e^{-E/(k_{B}T)}}.
\end{equation}
The equations for $\langle A_{n}\delta(w\pm
2\sqrt{2}n-qv_{cm})\rangle$ are  of the same form.
Finally, by simplifying and making a change of
variables in the integrals that represent the
expressions in brackets we get the following expression
for the response function:

\begin{eqnarray}
R(q,w) & = & \frac{2}{\pi^{2} q}
\frac{e^{-w^{2}/(q^{2}k_{B}T)}}{\sqrt{\pi k_{B}T}}
\int_{0}^{\pi/2}dz \int_{0}^{\pi}dy \nonumber \\
       &   & \times\  \{\exp[-\frac{k_{B}T}{8} q^{2}
(cos(y)-cos(y+z))^{2}] + \exp[-\frac{k_{B}T}{8} q^{2}
(cos(y)+cos(y+z))^{2}]\} \nonumber \\        &   &
\mbox{} + \frac{2}{\pi^{2} q \sqrt{\pi k_{B} T}}
\sum_{n=1}^{\infty}\{[e^{-(\frac{w+2\sqrt{2}n}{q})^{2}/
(k_{B}T)} + e^{-(\frac{w-2\sqrt{2}n}{q})^{2}/(k_{B}T)}]
\int_{0}^{\pi/2}dz \int_{0}^{\pi}dy \nonumber \\
       &   & \times\  (\exp[-\frac{k_{B}T}{8} q^{2}
(cos(y)-cos(y+z))^{2}] + \exp[-\frac{k_{B}T}{8} q^{2}
(cos(y)+cos(y+z))^{2}]) \nonumber \\        &   &
\times\  cos(2nz)\}.
\end{eqnarray}
This is our final expression for the response function of
a single two-quark hadron. It looks complicated, but it
is straightforward to calculate numerically
since we have a double integral of a smooth function.

\begin{figure}
1a) Value of the variational parameter $\lambda$\ which minimizes the energy
for
a given density $\rho$.  Results are for a quantum variational Monte Carlo
calculation involving 8 quarks using the wave function in Eq. (12).

1b) Trajectories for 8 quarks.

1c) Energy conservation in the algorithm used for the integration
of Newton's equation of motion. This is the total energy for 8
quarks as function of time. \end{figure}
\begin{figure}
2. Radial distribution function for $\rho = 0.414$ calculated in
two different ways. The dots and the stars represent a Metropolis
and a Molecular Dynamics calculation, respectively.
\end{figure}
\begin{figure}
3. The autocorrelation function versus time. The curve with the
stars corresponds to the free hadron case, the curve with the dots
and the error bars corresponds to the string flip model and the
solid curve correspond to the free quark case. The values of
momentum transfer and density are indicated at  the top of the
graph.
\end{figure}
\begin{figure}
4. The response function $R(q,w)$ as function of frequency $w$ for
a momentum transfer of $q = 0.034$ and a density of $\rho = 0.043$
. The curve with the stars corresponds to the free hadron
calculation, the curve with dots and error bars corresponds to the
string flip potential and the solid curve corresponds to the free
quark case.
\end{figure}
\begin{figure}
5. The same as in Fig. 4 but with $q = 0.33$ and $\rho = 0.212$ .
\end{figure}
\begin{figure}
6. The response function $R(q,w)$ with a fixed value of density
equal to $\rho = 0.414$ and three different values for the
momentum transfer $q$ .
 a) The same description as in Fig.4 but with $q = 0.32$ and $\rho
= 0.414$ .
 b) The same description as in Fig.4 but with $q = 0.65$ and $\rho
= 0.414$ .
 c) The same description as in Fig.4 but with $q = 1.30$ and $\rho
= 0.414$ .
\end{figure}
\begin{figure}
7. The same as in Fig.4 but with $q = 0.70$ and $\rho = 0.893$ .
\end{figure}
\begin{figure}
8. The same as in Fig.4 but with $q = 1.25$ and $\rho = 1.59$ .
\end{figure}
\begin{figure}
9. Finite size effects in the response function $R(q,w)$ for $q =
0.17$ and     $\rho = 0.21$ . The curve with the dots corresponds
to a calculation with 8 quarks and the curve with the stars
corresponds to a calculation with 16 quarks. \end{figure}
\begin{figure}
10. The same as in Fig.9 but with $q = 0.32$ and  $\rho = 0.414$.
\end{figure}

\end{document}